\documentclass[a4]{article}

\usepackage{graphics}
\usepackage{amssymb}

\begin{document}


\begin{center}
{\Huge Screening of hot gluon}\\
{\Large $\sim$ Lattice study of gluon screening masses $\sim$ }
\end{center}

\begin{center}
{A.~Nakamura, I.~Pushkina$^a$, T.~Saito and S.~Sakai$^b$}
\end{center}

\begin{center}{Research Institute for Information Science and Education,
 Hiroshima University,
 Higashi-Hiroshima 739-8521, Japan \\
$^a$School of Biosphere Sciences, Hiroshima University,
 Higashi-Hiroshima 739-8521, Japan \\
$^b$Faculty of Education, Yamagata University,
 Yamagata 990-8560, Japan }
\end{center}



\begin{abstract}
We calculate electric and magnetic masses of gluons between
$T=T_c$ and $6T_c$ using lattice QCD (quantum chromodynamics)
in the quench approximation for color $SU(3)$.
We find that
magnetic mass has finite values in this region, and the
temperature dependence of the electric
mass is consistent with that determined using the
hard-thermal-loop perturbation.
The hard-thermal-loop resummation improves significantly
the magnitude of the electric mass comparing to the leading
order perturbation.
Both screening masses have little gauge parameter dependence.
\end{abstract}




We are expecting relativistic heavy-ion collisions to
soon yield evidence of a quark-gluon plasma (QGP),
by creating sufficiently high temperature to bring the system
into the deconfining phase.
This new form of matter is a prediction of QCD and
will be studied for the first time in the laboratory.
One of the most urgent theoretical tasks is to investigate
the characteristics of gluons, the component of QGP
that controls the force, at a finite temperature based on QCD.
If we can calculate these features in a reliable manner,
phenomenological observations, such as the jet quenching,
may be analysed based on these knowledges.

There has been essential progress in the perturbative method
of treating hot QCD.
Using the hard-thermal-loop resummation technique\cite{BP},
we can now handle plasma effects of gauge theories.
The correct magnitude of the free energy from lattice QCD
was reproduced by such a calculation\cite{Andersen,Andersen2}.

Fundamental quantities of the hot gluon plasma
have been studied perturbatively.
Electric mass is given by the lowest perturbative calculation as
\begin{equation}
  m_{e,0} = \sqrt{\frac{N_c}{3}} g T .\label{ele}
\end{equation}
Magnetic mass, $m_m$,  is, however, beyond the scope of
the perturbative treatment\cite{Linde,Gross}.
It vanishes in the lowest perturbation, but if indeed
$m_m=0$, the gluons are essentially
unscreened, even if the electric mass is finite\cite{Niegawa}.
Also, zero magnetic mass causes serious infrared divergence and
makes consistent theoretical treatment very difficult
(see Chap.10.2 of Ref.\cite{Bellac}).

The maximum temperature accessible at RHIC or LHC is of the order of
$1\mbox{GeV}$ or $5T_c$
which is far below the regions where the perturbative calculation
 can be safely applied.
 Therefore nonperturbative analysis
is highly desirable in order to gain a quantitative understanding
of the properties of QGP.
Since the pioneering work 
for temperature dependencies of screening masses by Heller, Karsch and Rank
for SU(2) color QCD\cite{Heller}, there have been
several activities, but no $SU(3)$ lattice simulations of 
finite temperature gluon propagators in the literature.
Indeed, calculations of gauge-dependent observables are very
time-consuming and often unstable.

In this letter, we report the electric and magnetic screening masses
of gluons at finite temperature measured in lattice simulations
based on the SU(3) gauge theory, i.e., QCD in the quench approximation.
We calculate gluon propagators; which allows direct comparison with
perturbative calculations.

We calculate gluon correlation functions with finite momentum,
\begin{equation}
G_{\mu\nu}(P_t,P_x,P_y,z)=
\langle \mbox{Tr}
A_\mu(P_x,P_y,P_t,z)A_\nu(-P_x,-P_y,-P_t,0) \rangle\label{correl},
\end{equation}
where gauge potentials $A_\mu$'s are evaluated from the link variables
$U_\mu(x)\in$ SU(3) as
$A_\mu(x) = (U_\mu(x)-U_\mu^{\dag}(x))/2gi$.
The calculation of propagators with finite momenta enables us to
construct transverse and longitudinal components
without ambiguity\cite{Nakamura};
we define electric and magnetic parts of the gluon propagators as
\begin{equation}
\begin{array}{lll}
G_e(P,z)&=&\frac{Z}{2} \left[
      G_{tt}(\frac{2\pi}{N_x},0,0,z) +
      G_{tt}(0,\frac{2\pi}{N_y},0,z) \right]
,\\
G_m(P,z)&=&\frac{Z}{2} \left[
       G_{xx}(0,\frac{2\pi}{N_y},0,z) +
       G_{yy}(\frac{2\pi}{N_x},0,0,z) \right],
\label{Gem}
\end{array}
\end{equation}
which correspond to standard projected propagators in perturbative
calculations.
Here $P_\mu$'s are lattice momenta, and
the size of the lattice is given by $N_xN_yN_zN_t$.
At a sufficiently large distance, they are expected to behave as
\begin{equation}
G_{e(m)}(z) \propto e^{-E_{e(m)}z}\label{exp}.
\end{equation}

In order to quantize the theory and to extract gauge-dependent
observables,
we employ stochastic quantization $\grave{\mbox{a}}$
 la Zwanziger\cite{Zwanziger},
\begin{equation}
\frac{dA_{\mu}^a}{d\tau}=
-\frac{\delta S}{\delta A_{\mu}^a } +
\frac{1}{\alpha} D_{\mu}(A)^{ab}
\partial_{\nu} A_{\nu}^b + \eta_{\mu}^a\label{sq},
\label{SQeq}
\end{equation}
where $\tau$ is the Langevin time, the second term of
the r.h.s. is the gauge fixing term, and
$\eta$ represents Gaussian  random noise.
The gauge parameter is given by $\alpha$, and
$\alpha=0$ corresponds to the Lorentz gauge.
The lattice version of the algorithm was proposed in
Ref.\cite{Mizutani} as
\begin{equation}
U_{\mu}(x,\tau+\Delta\tau) =
\omega^{\dagger}(x,\tau)e^{if_{\mu}^a t^a}
U_{\mu}(x,\tau) \omega(x+\hat{\mu},\tau),
\label{LatSQeq}
\end{equation}
Here, $\omega$ means a gauge rotation matrix and
$f$ a driving force.
\begin{equation}
f_{\mu}^a=-\frac{\partial S}{\partial A_{\mu}^a} \Delta \tau
+ \eta^a \sqrt{\Delta \tau},
\hspace{1cm}
\omega = e^{i\beta \Delta^a \tau^a \Delta \tau /\alpha }.
\end{equation}
The gauge rotation step and Langevin step are executed alternately.

There are two reasons for using the stochastic gauge fixing method here:
one is practical and the other is conceptual.
Contrary to the standard Wilson-Mandula gauge fixing method\cite{Mandula},
where an iterative procedure is applied
and the number of iterations
is unpredictable for large lattices, in algorithm (\ref{LatSQeq}),
we repeat the steps of update and gauge rotation one after the other.
There is no convergence problem of gauge fixing.
Therefore we can precisely estimate CPU time.
Moreover, this algorithm, which may change
the gauge parameter $\alpha$,
has an advantage in testing gauge independence.
A conceptual problem is the Gribov ambiguity appearing in
non-Abelian gauge theories\cite{Gribov}.
Although our algorithm may not eliminate copies completely,
Zwanziger's stochastic gauge fixing algorithm maintains the gauge system
inside the Gribov region\cite{Zwanziger,Seiler}
if we start from the trivial configuration $\{A_{\mu}=0\}$.

We perform simulations on an
$N_x N_y N_z N_t =20\times 20 \times 32 \times 6$ lattice with
the standard Wilson gauge action.
The physical temperature is tuned
by changing the lattice cut-off scale 1/$a$[GeV], i.e., gauge coupling.
The relation between gauge coupling and the lattice scale is
estimated using the data in Ref.\cite{QCD_TARO}, which covers the range
from $\beta=6/g^2= 6$ to $7.1$. The confinement/deconfinement transition
temperature is set to be 256 MeV\cite{Boyd}. Then we
study the regions of $T/T_c=1\sim 6$.
The Runge-Kutta algorithm is applied to reduce
the finite Langevin step $\Delta\tau$ dependence\cite{Batrouni}.
We perform the simulation for a set of parameters with
$\Delta\tau = 0.03 \sim 0.05$
and extrapolate all results linearly to $\Delta\tau=0$.
The gauge parameter $\alpha$ is fixed to be one,
except in the gauge dependence test.

\begin{figure}[h]
\scalebox{0.325}{
\includegraphics{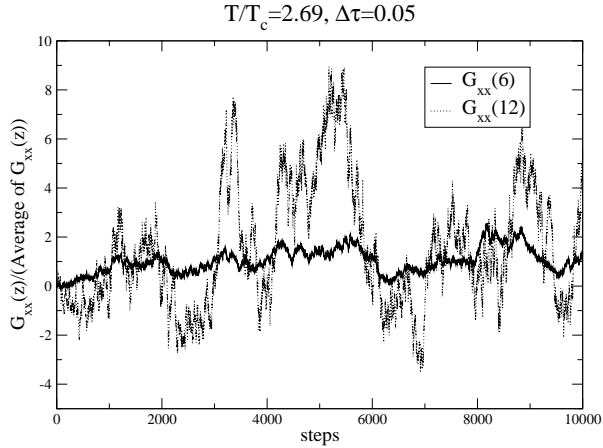}%
}
\caption{
Typical gluon behavior with Langevin steps.
We find that fluctuations of $G_{xx}(12)$ are much larger than
those of $G_{xx}(6)$.
In order to investigate the screening effect,
long-range contribution should be adopted.
Consequently we need a large number of statistics.
\label{hist1}}
\end{figure}

We observe a large fluctuation of gauge propagators, particularly
at large distances.  We show in Fig.\ref{hist1} a typical behavior
of gluon propagators $G(z)$ as a function of the Langevin step.
In order to estimate the necessary number of Langevin update steps,
we wait until $<G_{xx}(z)>$ and $<G_{yy}(z)>$ calculated in Eq.\ref{Gem}
becomes equal to each other.
A typical number of simulations is $0.2\sim 0.4$ million steps.

\begin{figure}[h]
\scalebox{0.325}{
\includegraphics{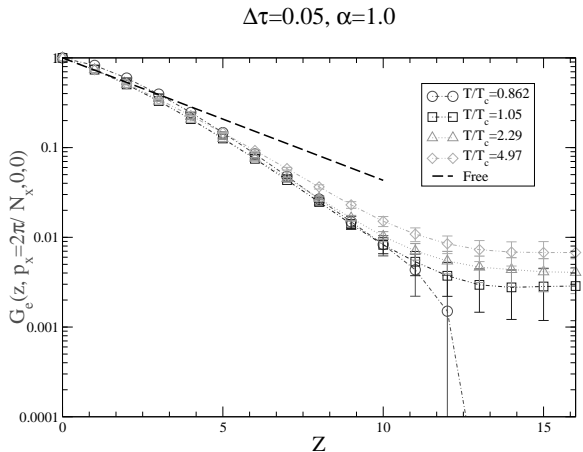} }
\caption{The electric propagator behavior in confinement (circle) and
deconfinement phase (the others)
with $P_x=\frac{2\pi}{N_x}$.
We find all propagators become massive 
comparing with the free propagator (long dashed line).
In confinement regions,
the propagator at long distance seems to have infinite mass
and vanishes, whereas
we find that the propagator beyond $T_c$, has finite mass.
\label{prop}}
\end{figure}

The dynamics of a gluon propagator is completely different
in confinement and deconfinement regions\cite{Nakamura2}.
In Fig.\ref{prop}, the gluon behavior is massive and
we see that
as distance $z$ increases, the mass of the propagator in confinement regions
seems to become infinite; the gluon is completely screened in the
confinement phase.
Consequently, we cannot employ the assumption, Eq.(\ref{exp}).
On the contrary, the propagator decreases
exponentially at a long distance with a finite damping rate
in the deconfinement region.

\begin{figure}[h]
\scalebox{0.45}{
\includegraphics{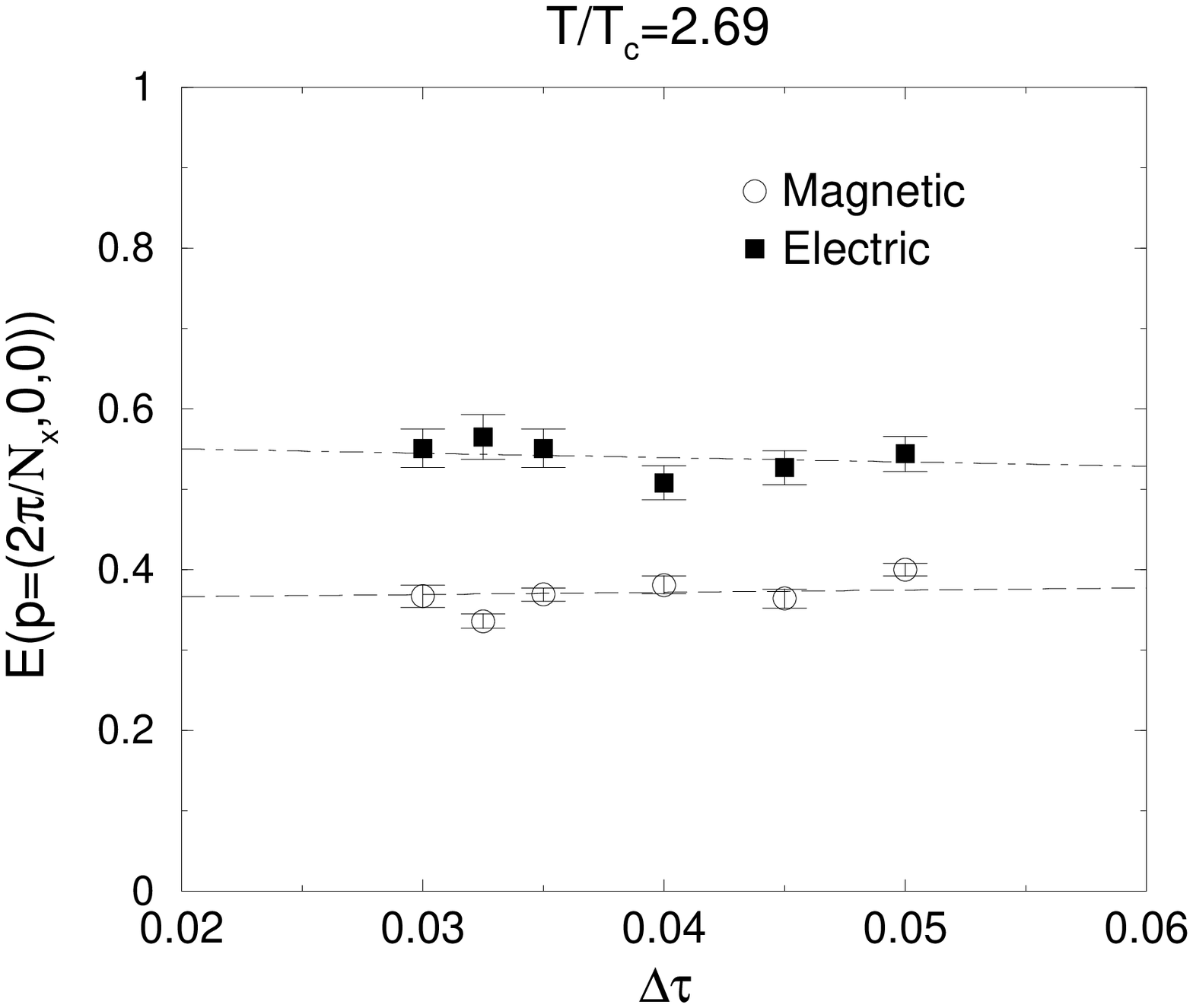} }
\caption{$\Delta \tau$ dependence of masses is slight.
Therefore, obtaining a final value at $\Delta \tau=0$,
we use the linear function for extrapolation.
Here, it is important to apply the Runge-Kutta algorithm
in order to suppress $\Delta \tau$ error.
\label{dtdeps}}
\end{figure}

To extract screening masses from the propagators
we use the following fitting functions
under the  lattice periodic condition:
\begin{equation}
 G_{e(m)} \sim \cosh(E_{e(m)}(p)(z-N_z/2)).
\end{equation}
We employ values for $z>1/T(=N_ta)$,
because
a screening effect occurs in a sufficiently long range.
Then, we obtain static mass from $E(p)$.

To get the final result at $\Delta \tau=0$,
we must extrapolate with respect to Langevin step width.
Fig.\ref{dtdeps} represents
$E(p)$ measured here versus $\Delta \tau$.
The slight dependences of $\Delta \tau$ cause us
to use a linear function when fitting data.

\begin{figure}
\scalebox{0.45}{
\includegraphics{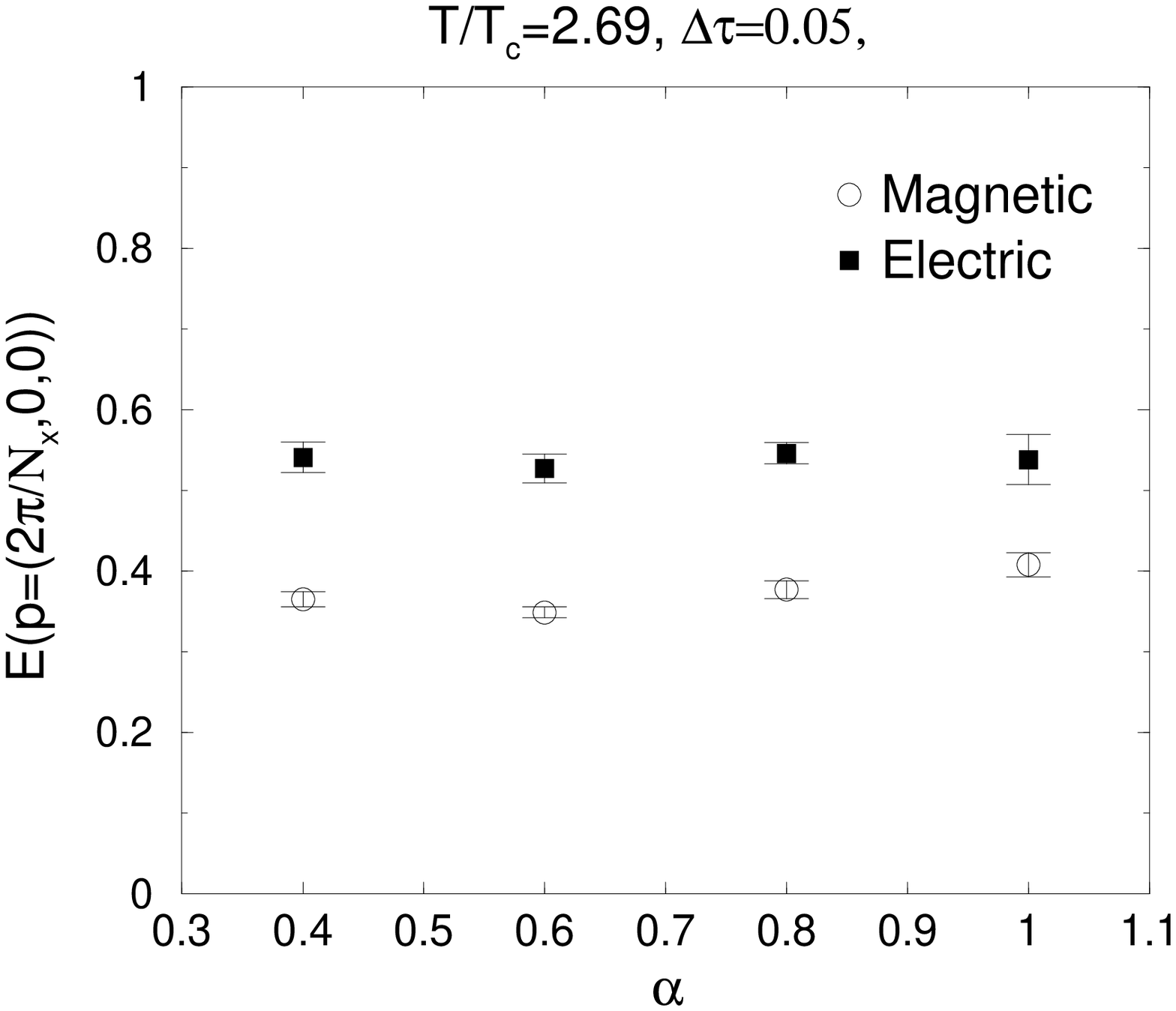}%
}
\caption{Gauge dependence graph for electric and magnetic
screening masses. Gauge dependences
of both screening masses are very slight in nonperturbative regions.
\label{meg}}
\end{figure}

Screening masses are physical and expected to be gauge independent.
However, since the definition of the gluon propagator,
Eq.(\ref{correl}) itself,
is generally gauge dependent,
it is nontrivial whether the screening masses
are gauge invariant or not.
In addition, since we fail to define the magnetic mass
by a perturbative analysis, it is particularly important to
check its gauge invariance.
In Fig.\ref{meg}, we show the gauge parameter $\alpha$ dependence of
electric and magnetic masses.  The result strongly suggests that
they are gauge independent and physical observables.

\begin{figure}
\scalebox{0.45}{
\includegraphics{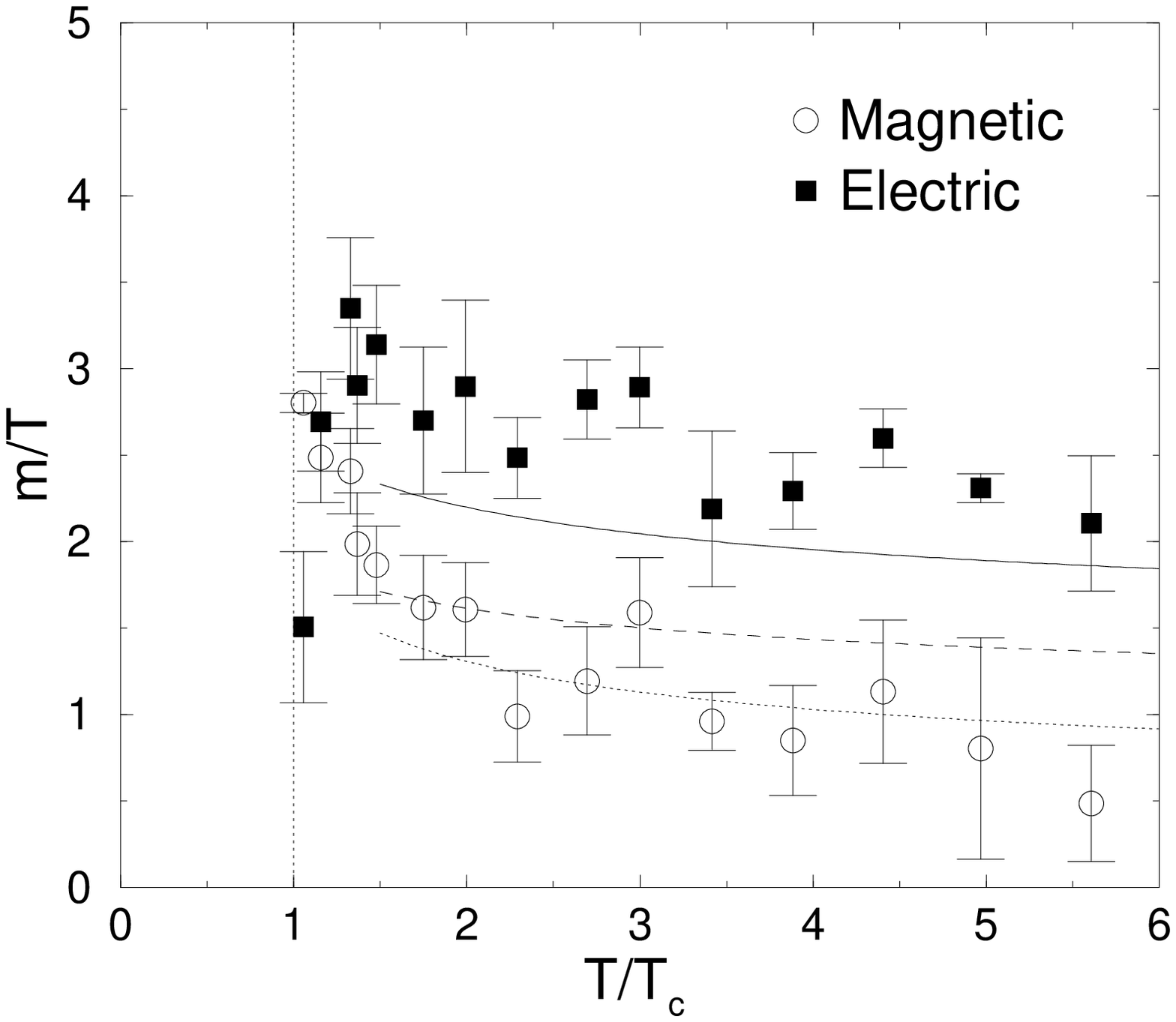} }
\caption{Temperature dependences
of electric and magnetic screening masses.
Dotted line was fitted based on the scaling of
$m_g\sim g^2T$.
For the electric mass,
the dashed and solid lines represent a leading order
perturbation and the hard-thermal-loop resummation result,
respectively.
\label{emt}}
\end{figure}

We study the temperature dependence of screening masses
at $T/T_c=1\sim6$ which would be expected to be realized
in high-energy heavy ion collision experiments such as RHIC or LHC.
Fig.3 shows the behavior of electric and magnetic
masses as a function of temperature.
The magnetic part definitely has nonzero mass in this
temperature region.
As $T$ increases, both masses decrease monotonically,
and at almost all temperatures, magnetic mass is less than electric mass,
except very near $T_c$ where the electric mass decreases very quickly
as $T$ approaches $T_c$.

We fit the data above $T \sim 1.5T_c$
by
\begin{equation}
\begin{array}{ll}
\displaystyle
\frac{m_e}{T}& = C_e g(T), \\
\displaystyle
\frac{m_m}{T}& = C_m g^2(T),\label{scale}
\end{array}
\end{equation}
which are predicted through the perturbative and
3-D reduction analysis\cite{Bellac}.
Here, we use running couplings as
\begin{equation}
g^2(\mu)=\frac{1}{2b_0\log(\mu/\Lambda)}
\left(1-\frac{b_1}{2b_0}\frac{\log(2\log\mu/\Lambda)}
{\log(\mu/\Lambda)} \right),
\end{equation}
and we set $\mu=2\pi T$ which is a Matsubara frequency, as the
renormalization point and $\Lambda = 1.03T_c$\cite{Andersen}
as the QCD mass scale.
$b_0=11N_c/48\pi^2$ and $b_1=(34/3)(N_c/(16\pi^2))^2$
are the first
two universal coefficients of the renormalization group.
We obtain $C_e=1.63(3)$, $\chi^2/\mbox{NDF}=0.715$
 and $C_m=0.482(31)$, $\chi^2/\mbox{NDF}=0.979$.
The scaling expected in Eq.(\ref{scale}) works well
for the electric mass although the magnitude $C_e$ is larger
than the leading order perturbation Eq.(\ref{ele}).

The hard-thermal-loop resummation technique
applying the free energy of hot gluon plasma
has been successful,
i.e., it gives the correct sign and roughly the correct
magnitude for $T>2T_c$\cite{Andersen}.
Rebhan gave a formula for the electric mass in the one-loop
resummed perturbation theory\cite{Rebhan},
\begin{equation}
\displaystyle
m_e^2=m_{e,0}^2
\left[ 1 + \frac{3g}{2\pi}\frac{m_e}{m_{e,0}}
\left( \log\frac{2m_e}{m_m} - \frac{1}{2} \right)+O(g^2)
 \right]
\end{equation}
Here, we assume the magnetic mass to be of the order of $g^2$.
Substituting our fitted value for $m_m$,
we can solve the above equation iteratively.
In Fig.3, we show this hard-thermal-loop resummation result together with
the lowest order calculation.
The hard-thermal-loop result gives a better description
 than naive perturbation,
upon comparing our numerical experiment.

In Ref.\cite{Alexanian}, the magnetic mass
was estimated as $m_g = 2.38N_c g^2 T/4\pi=0.568g^2 T$
using a self-consistent inclusion technique.
This is very close to our fitted result $C_m=0.515(35)$.

The electric mass was obtained also from
Polyakov line correlation functions at finite temperature
\cite{Gao,Kaczmarek} and those contradict our results.
3-D reduction argument\cite{Kajantie}
has shown that
$m_e/gT$ goes down when $T$ increases, but
even at $T\sim 1000\Lambda_{\overline{MS}}$
the electric mass is still about $3m_{e,0}$.

In conclusion, we have measured gluon propagators and obtained
the electric and magnetic masses
by lattice QCD simulations in the quench approximation for SU(3)
between $T=T_c$ and $6T_c$.
Features of QGP in this temperature region will be extensively studied
theoretically and experimentally in near future.
We observed that the magnetic mass does not vanish there;
To our knowledge, this is the first reliable measurement in SU(3).
It can be approximately fitted by $C_m g^2T$, and that
the electric mass is consistent with the hard-thermal-loop resummation
calculation. We also confirmed that
both electric and magnetic masses are gauge independent.
The calculational technique presented here can be applied to the study of
quark screening effects at finite temperature.

We would like to thank A. Ni$\acute{\mbox{e}}$gawa,
 S. Muroya and T. Inagaki for many
helpful discussions.  Numerical calculations were carried out on
SX5 at RCNP, Osaka Univ., VPP5000 at SIPC, Tsukuba Univ.
and HPC computer at INSAM, Hiroshima Univ..
This work is supported by Grants-in-Aid for Scientific Research from
Ministry of Education, Culture, Sports, Science and Technology,
Japan (No.11694085, No.11740159, and No. 12554008).


\label{}




\end{document}